\documentclass[runningheads]{llncs}
\usepackage{amsmath}
\usepackage{caption}
\usepackage{subcaption}
\usepackage{graphicx}
\usepackage{tikz}
\usetikzlibrary{backgrounds,decorations.pathreplacing,positioning}
\begin{document}
\title{PD-DWI: Predicting response to neoadjuvant chemotherapy in invasive breast cancer with  Physiologically-Decomposed Diffusion-Weighted MRI machine-learning model   
}
\titlerunning{Predicting response to neoadjuvant  chemotherapy in invasive breast cancer}
%
\author{Maya Gilad\inst{1}\orcidID{0000-0002-3869-5165} \and
Moti Freiman\inst{2}\orcidID{0000-0003-1083-1548}}
%
\authorrunning{M. Gilad and M. Freiman.}
%
\institute{Efi Arazi School of Computer Science, Reichman University, Herzliya, Israel, \and
Faculty of Biomedical Engineering, Technion, Haifa, Israel
\email{ms.maya.gilad@gmail.com}}
\maketitle              
\begin{abstract}
Early prediction of pathological complete response (pCR) following neoadjuvant chemotherapy (NAC) for breast cancer plays a critical role in surgical planning and optimizing treatment strategies. Recently, machine and deep-learning based methods were suggested for early pCR prediction from multi-parametric MRI (mp-MRI) data including dynamic contrast-enhanced MRI and diffusion-weighted MRI (DWI) with moderate success. We introduce PD-DWI, a physiologically decomposed DWI machine-learning model to predict pCR from DWI and clinical data. Our model first decomposes the raw DWI data into the various physiological cues that are influencing the DWI signal and then uses the decomposed data, in addition to clinical variables, as the input features of a radiomics-based XGBoost model.
We demonstrated the added-value of our PD-DWI model  over conventional machine-learning approaches for pCR prediction from mp-MRI data using the publicly available Breast Multi-parametric MRI for prediction of NAC Response (BMMR2) challenge. Our model substantially improves the area under the curve (AUC), compared to the current best result on the leaderboard (0.8849 vs. 0.8397) for the challenge test set. 
PD-DWI has the potential to improve prediction of pCR following NAC for breast cancer, reduce overall mp-MRI acquisition times and eliminate the need for contrast-agent injection.


\end{abstract}
\section{Introduction}
Breast cancer is the most prevalent cancer and is one of the leading causes of cancer mortality worldwide \cite{Banaie2018,Bhushan2021,Ferlay2021}. Neoadjuvant chemotherapy (NAC), a pre-operative standard-of-care for invasive breast cancer, has improved breast cancer treatment effectiveness and has been associated with better survival rates \cite{Bhushan2021,Song2021,Suo2021}. Pathological complete response (pCR), the absence of residual invasive disease in either breast or axillary lymph nodes, is used to assess patient response to NAC. While notable pCR rates have been reported even in the most aggressive tumors, approximately 20\% of breast cancers are resistant to NAC; hence, early response prediction of pCR might enable efficient selection and, when needed, alteration of treatment strategies and improve surgical planning \cite{Cain2019,Chen2020,Partridge2018,Song2021}. 
 
Multi-parametric MRI (mp-MRI) approaches were proposed to address the urgent need for early prediction of pCR.  
Commonly, dynamic contrast-enhanced magnetic resonance imaging (DCE-MRI) and diffusion-weighted MRI (DWI) are used in mp-MRI assessment of pCR due to their higher sensitivity to tissue architecture compared to anatomical MRI \cite{Baltzer2020,Gao2018,Liang2020,Partridge2018,Zhang2019}. Recently, radiomics and high-dimensional image features approaches were applied to predict pCR after NAC from mp-MRI data of breast cancer patients. Joo et al. \cite{joo2021multimodal} and Duanmu et al. \cite{duanmu2020prediction} introduced deep-learning models to fuse high‑dimensional mp-MRI image features, including DCE-MRI and DWI, and clinical information to predict pCR after NAC in  breast cancer. Huang et al. \cite{huang2021prediction}  and Liu et al. \cite{liu2019radiomics}  used radiomics features extracted from mp-MRI data for pCR prediction.

However, the mp-MRI data represents an overall aggregation of various physiological cues influencing the acquired signal. Specifically, DWI data with the computed apparent diffusion coefficient (ADC) map is known to represent an overall aggregation of both pure diffusion, inversely correlated with cellular density, and pseudo-diffusion, correlated with the collective motion of blood water molecules in the micro-capillary network \cite{Baltzer2020,Suo2021,Woodhams2005}.
As pCR is characterized by reduced cellular density, and therefore an increase in the presence of pure diffusion on one hand, and reduced blood flow in the capillaries and therefore reduced pseudo-diffusion on the other hand, the na\"ive aggregation of these cues by the ADC map may eliminate important features for pCR prediction \cite{Baltzer2020,Suo2021}. While separation between the diffusion and pseudo-diffusion components can be achieved by acquiring DWI data with multiple b-values and fitting the ``Intra-voxel incoherent motion'' (IVIM) model to the DWI data \cite{Baltzer2020,Liang2020,Suo2021}, lengthy acquisition times preclude clinical utilization \cite{Baltzer2020,Freiman2012}.

In this work, we present PD-DWI, a physiologically decomposed DWI model to predict pCR solely from clinical DWI data. 
Our model first decomposes the DWI data into pseudo-diffusion ADC and the pseudo-diffusion fraction maps representing the different physiological cues influencing the DWI signal using linear approximation. Then, we use these maps, in addition to clinical variables, as the input features of a radiomics-based XGBoost model for pCR prediction.

We demonstrated the added-value of our PD-DWI model over conventional machine-learning approaches for pCR prediction from mp-MRI data using the publicly available ``Breast Multiparametric MRI for prediction of NAC Response'' (BMMR2) challenge\footnote{\url{https://qin-challenge-acrin.centralus.cloudapp.azure.com/competitions/2}}. Our model substantially improves the area under the curve (AUC) compared to current best result on the leaderboard (0.8849 vs. 0.8397) on the challenge test set. 

Our approach does not require lengthy DWI data acquisition with multiple b-values and eliminates the need for Gadolinium-based contrast agent injection and DCE-MRI acquisition for early pCR prediction following NAC for invasive breast cancer patients.

\section{Method}
\subsection{Patient Cohort}
We used the BMMR2 challenge dataset. This dataset was curated from the ACRIN 6698 multi-center study \cite{Clark2013,Newitt2021,Partridge2018}. The dataset includes 191 subjects from multiple institutions, and has been pre-divided into training and test sub-groups (60\%-40\% split, stratified by pCR outcome) by challenge organizers. 

All subjects had longitudinal mp-MRI studies, including standardized DWI and DCE-MRI scans, at three time points: T0 (pre-NAC), T1 (3 weeks NAC), and T2 (12 weeks NAC). DWI data was acquired with a single-shot echo planar imaging sequence with parallel imaging (reduction factor, two or greater); fat suppression; a repetition time of greater than 4000 msec; echo time minimum; flip angle, 90$^\circ$; field of view, 300–360 mm; acquired matrix, 128 $\times$ 128 to 192 $\times$ 192; in-plane resolution, 1.7–2.8 mm; section thickness, 4–5 mm; and imaging time, 5 or fewer minutes. Diffusion gradients were applied in three orthogonal directions by using diffusion weightings (b-values) of 0, 100, 600, and 800 s/mm$^2$. No respiratory triggering or other motion compensation methods were used \cite{Partridge2018}. 

DWI whole-tumor manual segmentation and DCE-MRI functional tumor segmentation were provided for each time point, in addition to overall ADC and Signal Enhancement Ratio (SER) maps. Non-imaging clinical data included demographic data (age, race), 4-level lesion type, 4-level hormonal receptor HR/HER2 status, 3-level tumor grade, and MRI measured longest diameter (cm) at T0. 

Reference histopathological pCR outcome (0/1) was defined as no residual invasive disease in either breast or axillary lymph nodes on the basis of postsurgical histopathologic examination performed by the ACRIN 6698 institutional pathologists who were blinded to mp-MRI data \cite{Partridge2018}. The reference pCR data was available to challenge participants only for training sub-groups.

\subsection{Physiological decomposition of clinical breast DWI}
In the framework of DWI, the random displacement of individual water molecules in the tissue results in signal attenuation in the presence of magnetic field encoding gradient pulses. This attenuation increases with the ADC and the degree of sensitization-to-diffusion of the MRI pulse sequence (b-value), taking the form of a mono-exponential decay with the b-value:
\begin{align}
     s_i=s_0\exp\left(-b_i ADC\right)
    \label{eq:ADC}
\end{align}
where $s_i$ is the signal at b-value $b_i$ and $s_0$ is the signal without sensitizing the diffusion gradients.
While diffusion and blood micro-circulation are two entirely different physical phenomena, randomness results from the collective motion of blood water molecules in the micro-capillary network that may be depicted as a ``pseudo-diffusion'' process that influences the DWI signal decay for low b-values (0-200 s/mm$^2$) \cite{Freiman2012}.

The overall DWI signal attenuation can be produced with the bi-exponential IVIM signal decay model proposed by Le-Bihan \cite{le1988separation}:
\begin{align}
     s_i=s_0 \left(F \exp\left(-b_i \left(D^*+D\right)\right)+\left(1-F\right)\exp\left(-b_i D\right) \right)
\end{align}
where $D$ is the pure diffusion coefficient, $D^*$ is the pseudo-diffusion coefficient and $F$ is the pseudo-diffusion fraction.

Direct physiological decomposition of the BMMR2 challenge DWI data, by fitting the IVIM model to the DWI data, is not possible due to the limited number of b-values used during the acquisition. 
Instead, we used a linear approximation of the pseudo-diffusion and pseudo-diffusion fraction components as follows.
We first calculated an ADC map reflecting mostly pseudo-diffusion (ADC$_{0-100}$) with Eq.~\ref{eq:ADC} by using only DWI data acquired at low b-values ($\leq 100$ s/mm$^2$), overall ADC map using the entire DWI data  (ADC$_{0-800}$), and a pseudo-diffusion-free ADC map (ADC$_{100-800}$) using only DWI data acquired at high b-values (i.e $\geq100$ s/mm$^2$). Then, we calculate the pseudo-diffusion fraction map ($F$) which represents the relative contribution of the pseudo-diffusion cue to the overall DWI signal \cite{gurney2018comparison}.

\begin{figure}[t]
    \centering
    \begin{subfigure}[b]{0.48\textwidth}
         \centering
         \includegraphics[width=\textwidth]{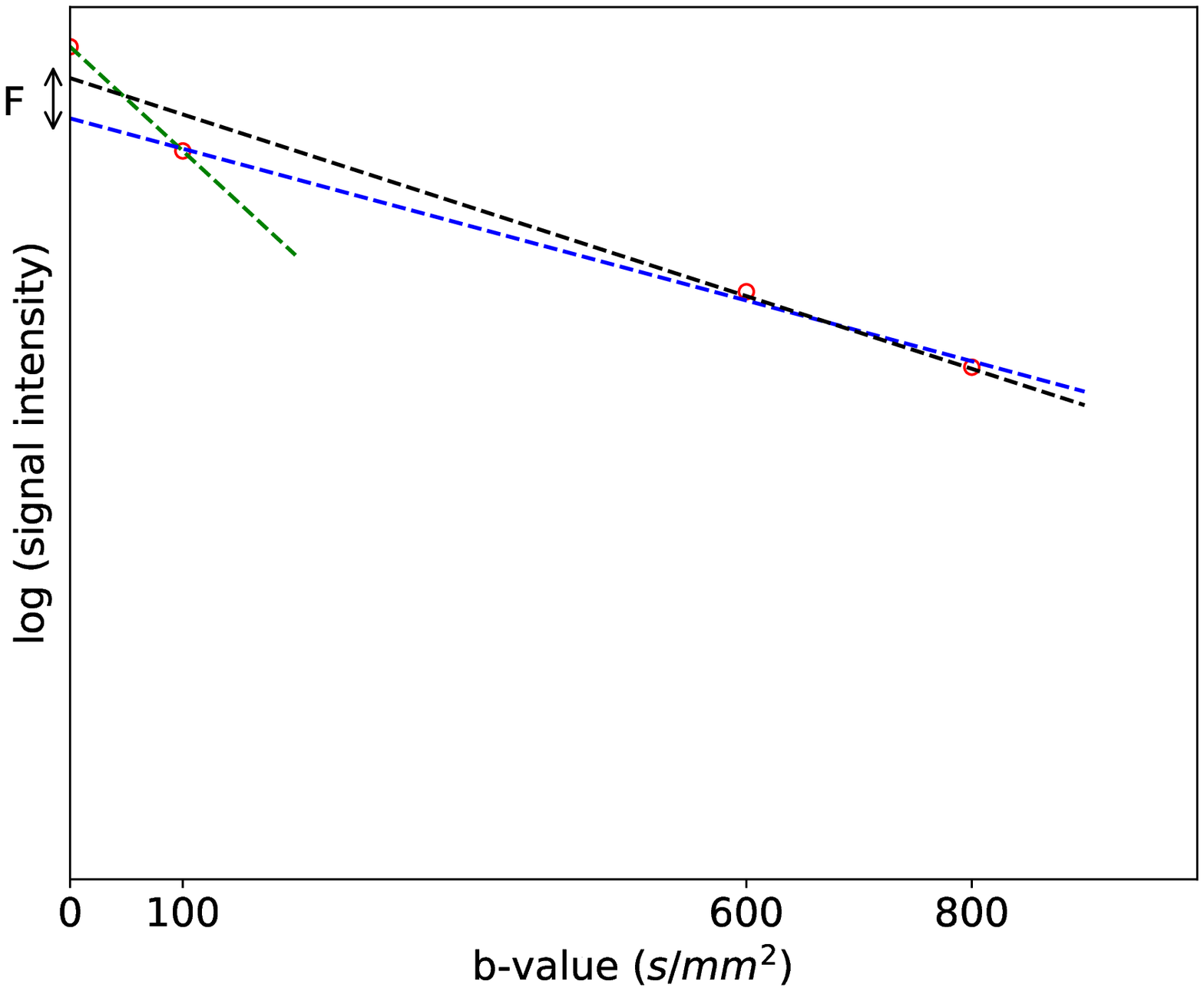}
         \caption{pCR}
    \end{subfigure}
     \begin{subfigure}[b]{0.48\textwidth}
         \centering
         \includegraphics[width=\textwidth]{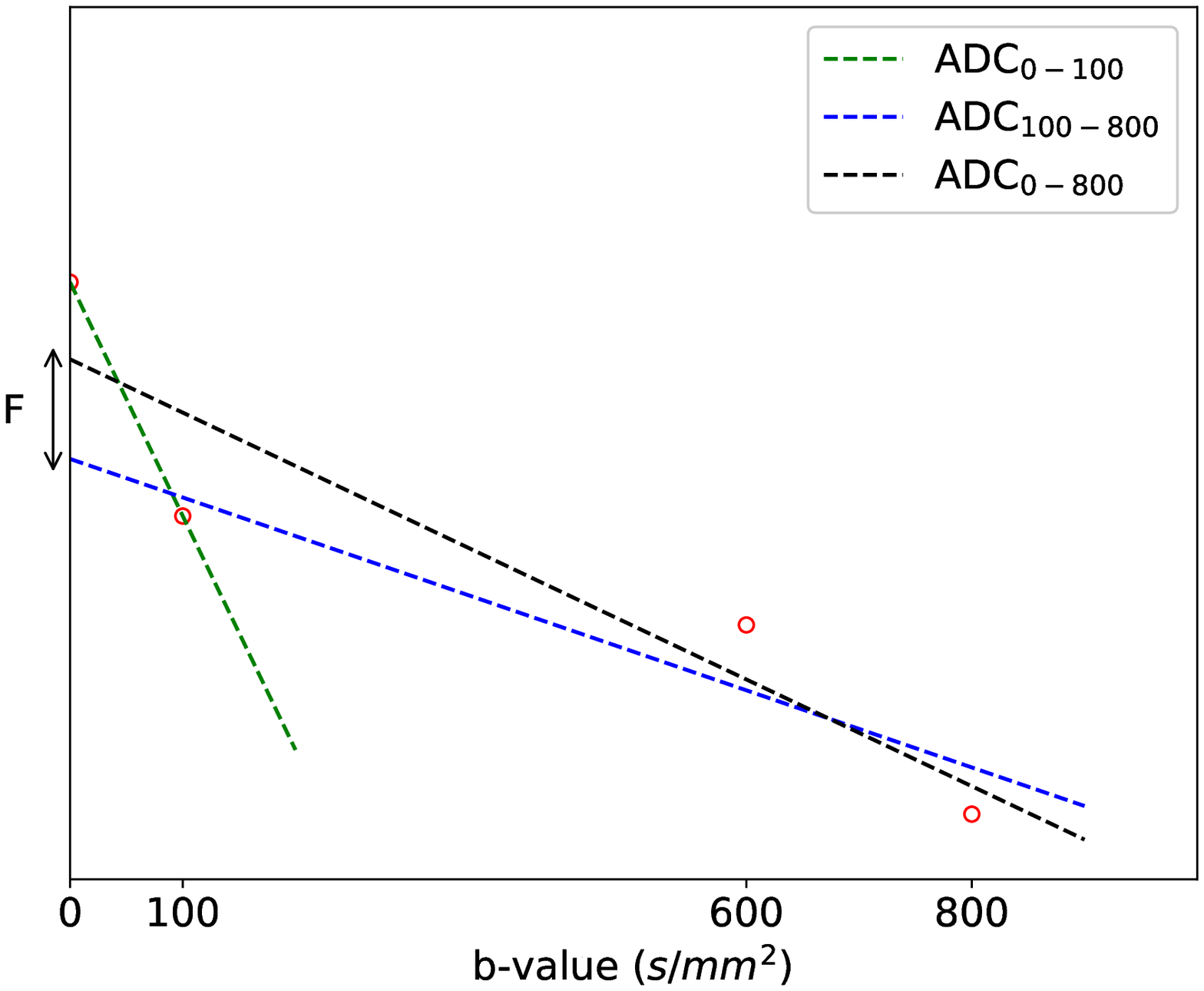}
         \caption{non-pCR}
    \end{subfigure}
       
    \caption{DWI signal decay as a function of the b-value along with the different ADC models of representative pCR and Non pCR patients.}
    \label{fig:adc}
\end{figure}

Fig.~\ref{fig:adc} presents the averaged DWI signal over the ROI for two representative patients from the BMMR2 challenge training set along with the ADC$_{0-100}$, ADC$_{0-800}$, ADC$_{100-800}$, and the pseudo-diffusion fraction (F) parameter data. The reduced slope of ADC$_{0-100}$ and F parameter data reflect reduced pseudo-diffusion and pseudo-diffusion fraction which represent a better pCR to NAC.

\begin{figure}[t]
    \centering
    \includegraphics[width=0.8\textwidth]{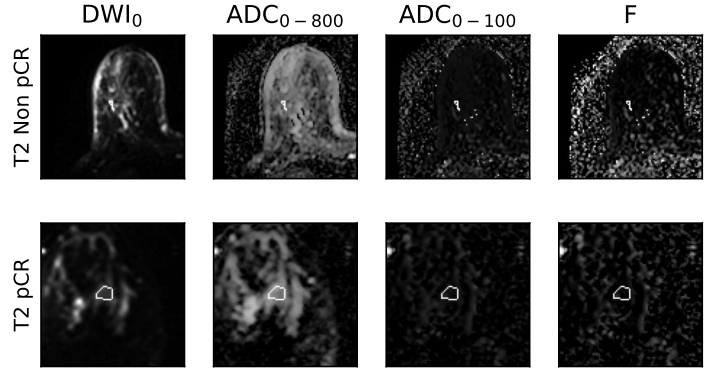}
    \caption{DWI data of pCR and non-PCR patients at T2 (mid-NAC). The white contour represents the tumor segmentation used for analysis as given by challenge organizers.}
    \label{fig:T2_MATRIX}
\end{figure}

Fig.~\ref{fig:T2_MATRIX} presents the DWI data along with the overall ADC map (ADC$_{0-800}$) and its physiological decomposition to the pseudo-diffusion ADC (ADC$_{0-100}$) and the pseudo-diffusion fraction (F).

\subsection{Machine Learning Model}

\begin{figure}
    \centering
    \begin{tikzpicture}[
    thumbnail/.style={rectangle, inner sep=0, draw=none, fill=none, minimum size=5mm},
    feature/.style={rectangle, inner sep=0, draw=black, fill=gray, minimum size=2.5mm},
    adc_feature/.style={feature, fill=blue},
    F_feature/.style={feature, fill=green},
    clinical_feature/.style={feature, fill=red},
    model/.style={rectangle, inner sep=2, draw=black,rounded corners},
    pcr/.style={circle, inner sep=2, draw=black}, scale = 0.75]
    ]
        
        \draw (0,0) -- (1,0) -- (1,2) -- (0, 2);
        
        \node[thumbnail] at (0,-0.5) (dwi0) [font=\footnotesize] {\begin{tabular}{c}\includegraphics[width=.1\linewidth]{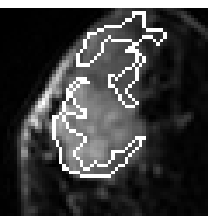} \\ DWI \end{tabular}};
        \node[thumbnail] at (0,2) (dwi1)  {\begin{tabular}{c}\includegraphics[width=.1\linewidth]{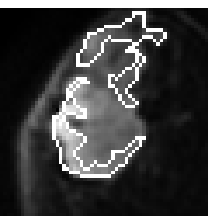} \end{tabular}};

        \draw[dotted] (dwi1.south) -- (dwi0.north) ;
        
        \draw [->] (3,1.1) -- (3,2.2) -- (4.5,2.2);
        \node[model,fill=white] at (3,1.1) (physics) {\begin{tabular}{c} Physiological \\ Feature Maps \\ Computations \end{tabular}};
        \draw [->] (1,1.1) -- (physics.west);
        
        \node[thumbnail]  (adc) at (5.5,-0.3) [font=\footnotesize] {\begin{tabular}{c} \includegraphics[width=.1\linewidth]{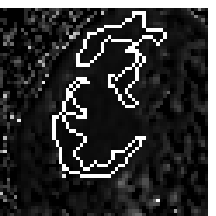} \\ ADC\textsubscript{0-100} \end{tabular}} ;
        \draw [->] (physics.south) |- (4.5,-0.3);
        
        \node[thumbnail] (F) at (5.5,1.9) [font=\footnotesize] {\begin{tabular}{c} \includegraphics[width=.1\linewidth]{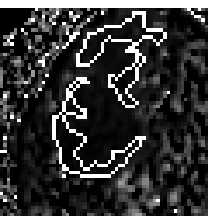} \\ F \end{tabular}} ;
        
        \draw [decorate,decoration={brace,amplitude=10},thick, black] (-0.75,3.2) -- (6.2,3.2) node [pos=0.5,above=7pt,font=\footnotesize] {\begin{tabular}{c} For all time points \end{tabular}};
        
        \node[adc_feature] at (7.8,-1) (r0) {};
        \node[adc_feature] at (7.8,-0.4) (r1) {};
        \node[adc_feature] at (7.8,0.1) (r2) {};
        \node[adc_feature] at (7.8,0.7)  (r3) {};
        \draw [dotted] (r0.north) -- (r1.south);
        \draw [dotted] (r1.north) -- (r2.south);
        \draw [dotted] (r2.north) -- (r3.south);
        \draw [->] (adc.east) -- (r0.west);
        \draw [->] (adc.east) -- (r1.west);
        \draw [->] (adc.east) -- (r2.west);
        \draw [->] (adc.east) -- (r3.west);
        
        \node[F_feature] at (7.8,1.5) (r4) {};
        \node[F_feature] at (7.8,1.95) (r5) {};
        \node[F_feature] at (7.8,2.55) (r6) {};
        \node[F_feature] at (7.8,3.2) (r7) {};
        \draw [dotted] (r4.north) -- (r5.south);
        \draw [dotted] (r5.north) -- (r6.south);
        \draw [dotted] (r6.north) -- (r7.south);
        \draw [->] (F.east) -- (r4.west);
        \draw [->] (F.east) -- (r5.west);
        \draw [->] (F.east) -- (r6.west);
        \draw [->] (F.east) -- (r7.west);
        
        \draw [decorate,decoration={brace,mirror,amplitude=10},thick, black] (6.2,-1.35) -- (7.4,-1.35) node [pos=0.5,below=7pt,font=\footnotesize] {\begin{tabular}{c} Radiomic \\ feature \\ extraction \end{tabular}};

        \node[clinical_feature] at (7.8,3.7) (c0) {};
        \node[clinical_feature] at (7.8,4.1) (c1) {};
        \node[clinical_feature] at (7.8,4.5) (c2) {};
        \node[clinical_feature] at (7.8,5.2) (c3) {};
        \draw [dotted] (c0.north) -- (c1.south);
        \draw [dotted] (c1.north) -- (c2.south);
        \draw [dotted] (c2.north) -- (c3.south);
        
        \draw [decorate,decoration={brace,mirror,amplitude=10},thick, black] (c3.north west) -- (c0.south west) node [pos=0.5,left=5pt,font=\footnotesize] {\begin{tabular}{c} Clinical \\ features \end{tabular}};
        
        \node[feature] at (9.2,0.25) (f0) {};
        \node[feature] at (9.2,1.25) (f1) {};
        \node[feature] at (9.2,1.75) (f2) {};
        \node[feature] at (9.2,2.25) (f3) {};
        \draw [dotted] (f0.north) -- (f1.south);
        \draw [dotted] (f1.north) -- (f2.south);
        \draw [dotted] (f2.north) -- (f3.south);
        
        \draw[->] (r1.east) -- (f0.west);
        \draw[->] (r2.east) -- (f1.west);
        \draw[->] (r6.east) -- (f2.west);
        \draw[->] (c1.east) -- (f3.west);
        
        \draw [decorate,decoration={brace,mirror,amplitude=10},thick, black] (8,-1.25) -- (9.2,-1.25) node [pos=0.5,below=7pt,font=\footnotesize] {\begin{tabular}{c} Feature \\ selection \end{tabular}};
        
        \node[model,fill=white] (cls) at (11,1.1) {\begin{tabular}{c} XGBoost \\ Classifier \end{tabular}};
        \draw [->] (f0.east) -- (cls.west);
        \draw [->] (f1.east) -- (cls.west);
        \draw [->] (f2.east) -- (cls.west);
        \draw [->] (f3.east) -- (cls.west);
        
    \end{tikzpicture}
    \caption{Model architecture:  The acquired DWI data at each time-point is decomposed into the different physiological cues. 3D Radiomics features are  extracted from the segmented tumor in ADC and F maps and categorical clinical data are concatenated. Then a feature selection process is applied to select the most informative features. Finally, the selected features are fed into an XGBoost classifier.}
    \label{fig:model}
\end{figure}

Our PD-DWI model was composed of physiological maps extracted from the DWI data in combination with clinical data.
Fig.~\ref{fig:model} presents the overall model architecture. The acquired DWI data at each time-point is decomposed into the different physiological cues reflecting pseudo-diffusion (ADC$_{0-100}$) and pseudo-diffusion-fraction (F). 3D Radiomics features were extracted from the segmented tumor in ADC and F maps using the PyRadiomics software package \cite{Griethuysen2017}.

We integrated clinical data in our model using several methods. By modeling the 4-level hormonal receptor status as two binary features (HR status and HER2 status) we allowed the model to leverage differences in radiomic features between tumor types. By converting the 3-level tumor grade to an ordinal categorical arrangement (Low - 1, Intermediate - 2, and High - 3) we allowed the model to use the relation between different severity levels. One patient in the training set lacked a tumor grade and was imputed as High using Most-Frequent imputation strategy. We applied One-Hot-Encoding on race and lesion type data.

The total number of extracted radiomic and clinical features was relatively high in comparison to the BMMR2 challenge dataset size. Thus, a subset of 100 features with the highest ANOVA F-values was selected in order to minimize model over-fitting and increase model robustness. Ultimately, an XGBoost classifier was trained on selected features and predicted the pCR probability of a given patient.

\subsection{Evaluation methodology}

\subsubsection{Relation between DWI signal attenuation decay and pCR prediction} 
We first explored the relation between signal attenuation decay and pCR prediction. We calculated additional ADC maps using various subsets of the b-values. Specifically, we calculated the additional following maps: ADC\textsubscript{100-800} using b-values 100-800,  ADC\textsubscript{0-800} using all available b-values, and finally ADC\textsubscript{0-100} using b-values 0 and 100. 

We assessed the impact of each component in our model by creating four additional models as follows. Three models were created using ADC\textsubscript{0-100}, ADC\textsubscript{100-800} and ADC\textsubscript{0-800}, without the F map at all. An additional model was created with only the F map. 

We assessed the performance of our PD-DWI model in comparison to the other DWI-based models based on the AUC metric achieved on the BMMR2 challenge test set using all available time-points (i.e. pre-NAC, early-NAC, and mid-NAC). The AUC metrics were calculated by BMMR2 challenge organizers upon submission of pCR predictions for all patients in the test set using the different models under evaluation as pCR outcomes of test set are not available to challenge participants.

\subsubsection{Early prediction of pCR over the course of NAC} 

We determined the added-value of our PD-DWI model in predicting pCR during the course of NAC in comparison to a baseline model that was built using the same pipeline, but include all imaging data available as part of the BMMR2 challenge dataset (ADC$_{0-800}$+SER). Specifically, the baseline model used both an ADC$_{0-800}$ map computed from the DWI data, and a SER map computed from the DCE-MRI data by the challenge organizers. 

We used the AUC computed by the challenge organizers on the test test as the evaluation metric. 

\subsubsection{Implementation details} 
All models were implemented\footnote{\url{https://github.com/TechnionComputationalMRILab/PD-DWI}} with python 3.8 and xgboost 1.5.1, scikit-learn 1.0.2, pandas 1.3.5 and pyradiomics 3.0.1 packages. All models were trained on the BMMR2 challenge training set. We removed one patient from the training set due to missing DWI image with b-value=100 s/mm$^2$. 

We performed hyper-tuning of number of features \textit{k} that were be fed to the XGBoost classifier. For XGBoost classifier, parameters hyper-tuning was focused on  \textit{min-child-weight},\textit{max-depth}, and \textit{subsample} parameters. By assigning  \textit{min-child-weight} with values larger than 1 (default) and limiting \textit{max-depth}, we guaranteed that nodes in boosted tree are only created when they represent a significant share of the samples and also limited the number of nodes in the boosted tree. The combination of both assures the boosted tree is based on a low number of features. Using the \textit{subsample} parameter we added another layer of mitigation against model-overfit by randomly selecting a new subset of samples on each training iteration. To avoid over-fitting, models were hyper-tuned on the training set using a K-fold cross-validation strategy.

We handled label imbalance (70\% non-pCR, 30\% pCR) using XGBoost's \textit{scale-pos-weight} parameter. We encouraged XGBoost to correct errors on pCR samples despite their smaller share in dataset by defining \textit{scale-pos-weight} as:
\begin{align}
     \frac{sum(\textit{non-pCR instances})}{sum(\textit{pCR instances})}
\end{align}

\section{Results}
\subsubsection{Relation between DWI signal attenuation decay and pCR prediction}
Table~\ref{tab:modelsettings} summarizes the comparison between all five model settings. Our PD-DWI model achieved higher AUC score than the ADC-only models and the F-only model. This result suggests changes in both cellular density and blood flow in the micro-capillary network as a response to NAC. Amongst the ADC-only models, ADC\textsubscript{0-800} had the highest AUC score, followed by ADC\textsubscript{0-100}. AUC scores of ADC\textsubscript{100-800} and F-only model scores were relatively similar.
Upon the release of test set labels, we evaluated the statistical significance of PD-DWI model compared to ADC-only models and the F-only model. PD-DWI model achieved the best Cohen's Kappa ($\kappa$) score whilst ADC\textsubscript{0-800} was second to last. PD-DWI model improved performance was statistically significant compared to F-only model (Sensitivity test, 0.65, $p<0.05$) and baseline model (Sensitivity test, 0.6, $p<0.05$).

\begin{table}[t]
\parbox{.45\linewidth}{
\centering
\begin{tabular}{c|c|c|c|}
\cline{2-4}
                                                 & AUC & $F_{1}$ & $\kappa$             \\ \hline
\multicolumn{1}{|c|}{Baseline}   & 0.8065 & 0.5714 & 0.4038\\ \hline
\multicolumn{1}{|c|}{ADC\textsubscript{0-100}-only}   & 0.8581 & 0.7391 & 0.6215\\ \hline
\multicolumn{1}{|c|}{ADC\textsubscript{100-800}-only}                   & 0.8465 & 0.6667 & 0.494          \\ \hline
\multicolumn{1}{|c|}{ADC\textsubscript{0-800}-only}   & 0.8781 & 0.6531 & 0.4823         \\ \hline
\multicolumn{1}{|c|}{F-only}   & 0.8423 & 0.6522 & 0.4953         \\ \hline
\multicolumn{1}{|c|}{PD-DWI}   & \textbf{0.8849} & \textbf{0.7391} & \textbf{0.6215}        \\ \hline
\end{tabular}
\caption{Models prediction performance comparison at time-point T2 (mid-NAC).}
\label{tab:modelsettings}
}
\hfill
\parbox{.45\linewidth}{
\centering
\begin{tabular}{c|c|}
\cline{2-2}
                        & AUC             \\ \hline
\multicolumn{1}{|c|}{Pre-challenge benchmark} & 0.78            \\ \hline
\multicolumn{1}{|c|}{DKFZ Team}               & 0.8031          \\ \hline
\multicolumn{1}{|c|}{IBM – BigMed}            & 0.8380          \\ \hline
\multicolumn{1}{|c|}{PennMed CBIG}            & 0.8397          \\ \hline
\multicolumn{1}{|c|}{PD-DWI}      & \textbf{0.8849} \\ \hline
\end{tabular}
\caption{BMMR2 Challenge performance\protect\footnotemark \text{ }comparison at time-point T2 (mid-NAC). }
\label{tab:challenge}
}
\end{table}
\footnotetext{\url{https://qin-challenge-acrin.centralus.cloudapp.azure.com/competitions/2\#results}}

Table~\ref{tab:challenge} presents a comparison between the results obtained by our PD-DWI model and the current BMMR2 challenge leaderboard at time-point T2 (mid-NAC). Our PD-DWI model outperformed the best score in the challenge. To the best of our knowledge, our model has the best pCR prediction performance for the BMMR2 dataset.

\subsubsection{Early prediction of pCR over the course of NAC} 
\begin{figure}[t]
    \centering
    \includegraphics[width=0.65\textwidth]{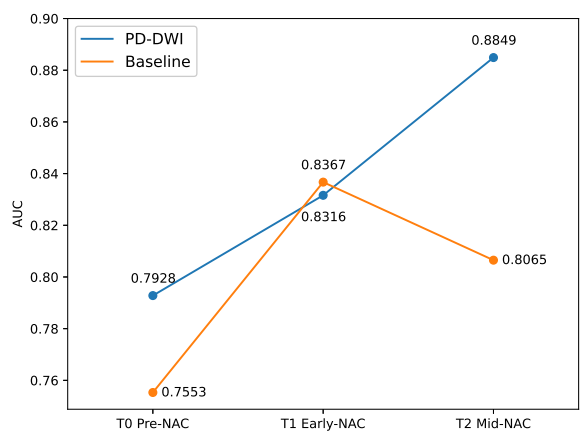}
    \caption{Model performance at the different phases of NAC treatment.}
    \label{fig:timepointscomparison}
\end{figure}

We evaluated the added-value of the PD-DWI model for early prediction of pCR following NAC compared to the baseline model. Fig.~\ref{fig:timepointscomparison} summarizes model performance as a function of the different time points of NAC treatment. As expected both models' pCR predictions improve given additional imaging data acquired during the course of the treatment. Specifically, our PD-DWI model consistently outperforms the baseline model in both T0 and T2. While our model performed about the same on T1, our approach does not require contrast injection and DCE-MRI acquisition. These findings suggest that changes in pseudo-diffusion and pseudo-diffusion fraction are more evident compared to changes in overall ADC values and SER values during the course of the treatment. Finally, our PD-DWI model performance using only pre-NAC data achieved a higher pCR prediction performance compared to the BMMR2 benchmark which used all time-points as provided by challenge organizers (0.7928 vs 0.78). 

\section {Conclusions}

We introduced PD-DWI, a physiologically decomposed DWI machine-learning model, to predict pCR following NAC in invasive breast cancer. Our model accounts for the different physiological cues associated with pCR as reflected by the DWI signal rather than using aggregated information by means of the ADC map. The proposed PD-DWI approach demonstrated a substantial improvement in pCR prediction over the best published results on a publicly available challenge for pCR prediction from mp-MRI data. 
Our model is based solely on clinical DWI data, thus eliminating the need for lengthy DWI acquisition times, Gadolinium-based contrast agent injections, and DCE-MRI imaging.
The proposed approach can be directly extended for the prediction of pCR following NAC in additional oncological applications.

\bibliographystyle{splncs04}
\bibliography{refs}

\begin{thebibliography}{10}
\providecommand{\url}[1]{\texttt{#1}}
\providecommand{\urlprefix}{URL }
\providecommand{\doi}[1]{https://doi.org/#1}

\bibitem{Baltzer2020}
Baltzer, P., Mann, R.M., Iima, M., Sigmund, E.E., Clauser, P., Gilbert, F.J.,
  Martincich, L., Partridge, S.C., Patterson, A., Pinker, K., et~al.:
  Diffusion-weighted imaging of the breast—a consensus and mission statement
  from the eusobi international breast diffusion-weighted imaging working
  group. European radiology  \textbf{30}(3),  1436--1450 (2020)

\bibitem{Banaie2018}
Banaie, M., Soltanian-Zadeh, H., Saligheh-Rad, H.R., Gity, M.: Spatiotemporal
  features of dce-mri for breast cancer diagnosis. Computer Methods and
  Programs in Biomedicine  \textbf{155} (2018).
  \doi{10.1016/j.cmpb.2017.12.015}

\bibitem{Bhushan2021}
Bhushan, A., Gonsalves, A., Menon, J.U.: Current state of breast cancer
  diagnosis, treatment, and theranostics. Pharmaceutics  \textbf{13} (2021).
  \doi{10.3390/pharmaceutics13050723}

\bibitem{Cain2019}
Cain, E.H., Saha, A., Harowicz, M.R., Marks, J.R., Marcom, P.K., Mazurowski,
  M.A.: Multivariate machine learning models for prediction of pathologic
  response to neoadjuvant therapy in breast cancer using mri features: a study
  using an independent validation set. Breast Cancer Research and Treatment
  \textbf{173} (2019). \doi{10.1007/s10549-018-4990-9}

\bibitem{Chen2020}
Chen, X., Chen, X., Yang, J., Li, Y., Fan, W., Yang, Z.: Combining dynamic
  contrast-enhanced magnetic resonance imaging and apparent diffusion
  coefficient maps for a radiomics nomogram to predict pathological complete
  response to neoadjuvant chemotherapy in breast cancer patients. Journal of
  computer assisted tomography  \textbf{44} (2020).
  \doi{10.1097/RCT.0000000000000978}

\bibitem{Clark2013}
Clark, K., Vendt, B., Smith, K., Freymann, J., Kirby, J., Koppel, P., Moore,
  S., Phillips, S., Maffitt, D., Pringle, M., Tarbox, L., Prior, F.: The cancer
  imaging archive (tcia): Maintaining and operating a public information
  repository (Jul 2013). \doi{10.1007/s10278-013-9622-7},
  \url{http://dx.doi.org/10.1007/s10278-013-9622-7}

\bibitem{duanmu2020prediction}
Duanmu, H., Huang, P.B., Brahmavar, S., Lin, S., Ren, T., Kong, J., Wang, F.,
  Duong, T.Q.: Prediction of pathological complete response to neoadjuvant
  chemotherapy in breast cancer using deep learning with integrative imaging,
  molecular and demographic data. In: International conference on medical image
  computing and computer-assisted intervention. pp. 242--252. Springer (2020)

\bibitem{Ferlay2021}
Ferlay, J., Colombet, M., Soerjomataram, I., Parkin, D.M., Piñeros, M., Znaor,
  A., Bray, F.: Cancer statistics for the year 2020: An overview. International
  Journal of Cancer  \textbf{149} (2021). \doi{10.1002/ijc.33588}

\bibitem{Freiman2012}
Freiman, M., Voss, S.D., Mulkern, R.V., Perez-Rossello, J.M., Callahan, M.J.,
  Warfield, S.K.: In vivo assessment of optimal b-value range for
  perfusion-insensitive apparent diffusion coefficient imaging. Medical Physics
   \textbf{39} (2012). \doi{10.1118/1.4736516}

\bibitem{Gao2018}
Gao, W., Guo, N., Dong, T.: Diffusion-weighted imaging in monitoring the
  pathological response to neoadjuvant chemotherapy in patients with breast
  cancer: A meta-analysis. World Journal of Surgical Oncology  \textbf{16}
  (2018). \doi{10.1186/s12957-018-1438-y}

\bibitem{Griethuysen2017}
Griethuysen, J.J.V., Fedorov, A., Parmar, C., Hosny, A., Aucoin, N., Narayan,
  V., Beets-Tan, R.G., Fillion-Robin, J.C., Pieper, S., Aerts, H.J.:
  Computational radiomics system to decode the radiographic phenotype. Cancer
  Research  \textbf{77} (2017). \doi{10.1158/0008-5472.CAN-17-0339}

\bibitem{gurney2018comparison}
Gurney-Champion, O.J., Klaassen, R., Froeling, M., Barbieri, S., Stoker, J.,
  Engelbrecht, M.R., Wilmink, J.W., Besselink, M.G., Bel, A., Van~Laarhoven,
  H.W., et~al.: Comparison of six fit algorithms for the intra-voxel incoherent
  motion model of diffusion-weighted magnetic resonance imaging data of
  pancreatic cancer patients. PloS one  \textbf{13}(4),  e0194590 (2018)

\bibitem{huang2021prediction}
Huang, Y., Chen, W., Zhang, X., He, S., Shao, N., Shi, H., Lin, Z., Wu, X., Li,
  T., Lin, H., et~al.: Prediction of tumor shrinkage pattern to neoadjuvant
  chemotherapy using a multiparametric mri-based machine learning model in
  patients with breast cancer. Frontiers in Bioengineering and Biotechnology
  p.~558 (2021)

\bibitem{joo2021multimodal}
Joo, S., Ko, E.S., Kwon, S., Jeon, E., Jung, H., Kim, J.Y., Chung, M.J., Im,
  Y.H.: Multimodal deep learning models for the prediction of pathologic
  response to neoadjuvant chemotherapy in breast cancer. Scientific reports
  \textbf{11}(1), ~1--8 (2021)

\bibitem{le1988separation}
Le~Bihan, D., Breton, E., Lallemand, D., Aubin, M., Vignaud, J., Laval-Jeantet,
  M.: Separation of diffusion and perfusion in intravoxel incoherent motion mr
  imaging. Radiology  \textbf{168}(2),  497--505 (1988)

\bibitem{Liang2020}
Liang, J., Zeng, S., Li, Z., Kong, Y., Meng, T., Zhou, C., Chen, J., Wu, Y.P.,
  He, N.: Intravoxel incoherent motion diffusion-weighted imaging for
  quantitative differentiation of breast tumors: A meta-analysis. Frontiers in
  Oncology  \textbf{10} (2020). \doi{10.3389/fonc.2020.585486}

\bibitem{liu2019radiomics}
Liu, Z., Li, Z., Qu, J., Zhang, R., Zhou, X., Li, L., Sun, K., Tang, Z., Jiang,
  H., Li, H., et~al.: Radiomics of multiparametric mri for pretreatment
  prediction of pathologic complete response to neoadjuvant chemotherapy in
  breast cancer: a multicenter study. Clinical Cancer Research
  \textbf{25}(12),  3538--3547 (2019)

\bibitem{Newitt2021}
Newitt, D.C., Partridge, S.C., Zhang, Z., Gibbs, J., Chenevert, T., Rosen, M.,
  Bolan, P., Marques, H., Romanoff, J., Cimino, L., Joe, B.N., Umphrey, H.,
  Ojeda-Fournier, H., Dogan, B., Oh, K.Y., Abe, H., Drukteinis, J., Esserman,
  L.J., Hylton, N.M.: Acrin 6698/i-spy2 breast dwi (2021).
  \doi{10.7937/TCIA.KK02-6D95},
  \url{https://wiki.cancerimagingarchive.net/x/lwH9Ag}

\bibitem{Partridge2018}
Partridge, S.C., Zhang, Z., Newitt, D.C., Gibbs, J.E., Chenevert, T.L., Rosen,
  M.A., Bolan, P.J., Marques, H.S., Romanoff, J., Cimino, L., Joe, B.N.,
  Umphrey, H.R., Ojeda-Fournier, H., Dogan, B., Oh, K., Abe, H., Drukteinis,
  J.S., Esserman, L.J., Hylton, N.M.: Diffusion-weighted mri findings predict
  pathologic response in neoadjuvant treatment of breast cancer: The acrin 6698
  multicenter trial. Radiology  \textbf{289} (2018).
  \doi{10.1148/radiol.2018180273},
  \url{https://pubs.rsna.org/doi/full/10.1148/radiol.2018180273}

\bibitem{Song2021}
Song, D., Man, X., Jin, M., Li, Q., Wang, H., Du, Y.: A decision-making
  supporting prediction method for breast cancer neoadjuvant chemotherapy.
  Frontiers in Oncology  \textbf{10} (2021). \doi{10.3389/fonc.2020.592556}

\bibitem{Suo2021}
Suo, S., Yin, Y., Geng, X., Zhang, D., Hua, J., Cheng, F., Chen, J., Zhuang,
  Z., Cao, M., Xu, J.: S. Journal of Translational Medicine  \textbf{19}
  (2021). \doi{10.1186/s12967-021-02886-3}

\bibitem{Woodhams2005}
Woodhams, R., Matsunaga, K., Kan, S., Hata, H., Ozaki, M., Iwabuchi, K.,
  Kuranami, M., Watanabe, M., Hayakawa, K.: Adc mapping of benign and malignant
  breast tumors. Magnetic Resonance in Medical Sciences  \textbf{4} (2005).
  \doi{10.2463/mrms.4.35}

\bibitem{Zhang2019}
Zhang, M., Horvat, J.V., Bernard-Davila, B., Marino, M.A., Leithner, D.,
  Ochoa-Albiztegui, R.E., Helbich, T.H., Morris, E.A., Thakur, S., Pinker, K.:
  Multiparametric mri model with dynamic contrast-enhanced and
  diffusion-weighted imaging enables breast cancer diagnosis with high
  accuracy. Journal of Magnetic Resonance Imaging  \textbf{49} (2019).
  \doi{10.1002/jmri.26285}

\end{thebibliography}

\end{document}